\documentclass[iop,apj,tighten]{emulateapj}
\usepackage{amsmath,amstext}
\usepackage{amsmath,amssymb}
\usepackage{bm}
\usepackage[breaklinks,colorlinks,citecolor=blue,linkcolor=magenta]{hyperref} 
\usepackage[all]{hypcap} 
\usepackage[usenames,dvipsnames]{color}

\DeclareMathAccent{\dot}    {\mathalpha}{operators}{'137} 
\DeclareMathAccent{\ddot}    {\mathalpha}{operators}{'177} 
\shorttitle{Isotropic-Nematic Phase Transitions II}
\shortauthors{Tak\'acs \& Kocsis}

\begin{document}
\title{Isotropic-Nematic Phase Transitions in Gravitational Systems II: \\ Higher Order Multipoles}
\author{\'Ad\'am Tak\'acs and Bence Kocsis}
\affil{Institute of Physics, E\"otv\"os University, P\'azm\'any P. s. 1/A, Budapest, 1117, Hungary;}
\email{tadam@caesar.elte.hu}
\email{bkocsis@gmail.com}

\begin{abstract}
The gravitational interaction among bodies orbiting in a spherical potential leads to the rapid relaxation of the orbital planes' distribution, a process called vector resonant relaxation. We examine the statistical equilibrium of this process for a system of bodies with similar semimajor axes and eccentricities. We extend the previous model of \citet{Roupas+2017} by accounting for the multipole moments beyond the quadrupole, which dominate the interaction for radially overlapping orbits. Nevertheless, we find no qualitative differences between the behavior of the system with respect to the model restricted to the quadrupole interaction. The equilibrium distribution resembles a counterrotating disk at low temperature and a spherical structure at high temperature. The system exhibits a first order phase transition between the disk and the spherical phase in the canonical ensemble if the total angular momentum is below a critical value. We find that the phase transition erases the high order multipoles, i.e. small-scale structure in angular momentum space, most efficiently. The system admits a maximum entropy and a maximum energy, which lead to the existence of negative temperature equilibria.
\end{abstract}
\keywords{galaxies: evolution --- galaxies: nuclei --- galaxies: structure --- Galaxy: center --- Galaxy: nucleus --- stars: kinematics and dynamics}
\maketitle

\section{Introduction}\label{sec:introduction}
Most galaxies contain a supermassive black hole (SMBH) at their centers with mass $M_\bullet\approx10^6-10^{10} M_\odot$. In many cases the SMBH is surrounded by a very dense stellar system of a few parsecs called nuclear star cluster (NSC) \citep{2013ARA&A..51..511K}. Understanding the processes within the NSC may be key to understand the growth of SMBHs along with the large scale structure of the host galaxy which is regulated by the SMBH.

The dynamical evolution in such systems is governed by several processes operating on different timescales (see \citealt{Rauch:1996fb,Hopman:2006qr,Gurkan:2007bj,Eilon:2008ht,Kocsis+Tremaine2011,Kocsis+Tremaine2015}). At leading order bodies follow eccentric, near-Keplerian orbits with period $t_\text{orb}$. On longer timescales, $t_\text{aps}$, the Newtonian gravity of the spherical mass distribution and general relativistic effects lead to apsidal precession. On even longer timescales, $t_{\rm vrr}$, the non-spherical components of the gravitational field of the stellar system leads to diffusion in the orientation of the orbits without changing the magnitude of angular momenta and the mechanical (or binding) energy of the orbits around the SMBH, resulting in the conservation of the orbital eccentricities and semimajor axes. This process is called \textit{vector resonant relaxation} (VRR). On still longer timescales, $t_{\rm srr}$, the non-axisymmetric torques between the stellar orbits leads to a diffusion in the eccentricities (or equivalently the magnitudes of the angular momenta), a process called \textit{scalar resonant relaxation} (SRR). Due to the close encounters of the bodies, the two-body relaxation changes the semimajor axes or Keplerian energies of the orbits on the longest timescale. Therefore the dynamical evolution is strongly collisional at the $t_{\rm vrr}$ timescale and above. These timescales are estimated as \citet{Kocsis+Tremaine2011}
\begin{enumerate}
\item orbital period: $t_\text{orb}\sim(GM_\bullet/r^2)^{-1/2}\sim 1-10^4$ yr;
\item apsidal precession:\\$t_\text{aps}\sim t_\text{orb}M_\bullet/(Nm)\sim 10^3-10^5$ yr;
\item orbital plane reorientation:\\$t_{\rm vrr}\sim t_\text{orb}M_\bullet/(\sqrt{N}m)\sim 10^6-10^7$ yr;
\item angular momentum diffusion:\\$t_{\rm srr}\sim t_\text{orb}M_\bullet/m\sim 10^8-10^{10}$ yr;
\item Keplerian energy diffusion:\\$t_{\rm 2-body}\sim t_\text{orb}^2M_\bullet/(Nm^2)\sim 10^9-10^{10}$ yr.
\end{enumerate}
More generally, this hierarchy applies for collisional gravitational systems dominated by a point mass or spherical potential (e.g. planetary systems, moon systems, etc.). Globular clusters may also exhibit vector resonant relaxation \citep[in prep.]{MeironKocsis}.

The statistical physics of gravitational systems have been examined by several previous works \citep[see e.g.][and references therein]{2000ApJ...531..739N,Chavanis:2002rj,2005MNRAS.361..385A,2014PhR...535....1L,2014arXiv1401.5534T,0004-637X-807-2-157,0004-637X-820-2-129, 2016JPhA...49r5002R, 2016MNRAS.458.4129S, 2016MNRAS.458.4143S,2017MNRAS.465.1856S,2015A&A...584A.129F,2017MNRAS.471.2642F}. Recently, in a companion paper \citet{Roupas+2017} have examined the statistical equilibrium of orbital planes after the VRR process is completed for one component systems in which the semimajor axis and eccentricities are the same for all objects. They truncated the interaction at the quadrupole order in a multipole expansion. These simplifying assumptions have led to a tractable model in which the statistical equilibrium structure of phase space could be completely mapped out. The results showed that the stable distribution of bodies close to zero temperature represents an axisymmetric thin disk in which the bodies may orbit in both senses and where the disk thickness increases with temperature. Anisotropic (i.e. biaxial) long-lived metastable equilibria were also found, which consist of two disks with a relative inclination larger than 90 degrees. If the total angular momentum is smaller than a critical value, the system exhibits a first order phase transition in the canonical ensemble from the disk phase to a nearly spherical distribution. Furthermore the results showed that negative temperature equilibria are possible and are stable, a curious phenomenon in condensed matter physics \citep[see in][]{Braun52,2015AmJPh..83..163F,Dunkel:2013fha,PhysRevE.91.052147,2015JSMTE..12..002C,PhysRevE.93.032149}. 

This is the second paper in this series, where we extend \citet{Roupas+2017} to examine the statistical equilibrium of orbital planes due to VRR by relaxing the quadrupole-interaction approximation. \citet{Kocsis+Tremaine2015} have shown that the quadrupole approximation dominates the dynamics for radially widely separated orbits, but for systems with radially overlapping orbits\footnote{Radially overlapping refers to the case where the radial distance from the center for two Keplerian elliptical orbits between their respective peri- and apo-apsides have a non-empty intersection \citep{Kocsis+Tremaine2015}.} the contribution of higher $\ell$ multipoles dominates for mutual inclinations less than $\mathcal{O}(1/\ell)$. In this case, the Hamilton-equations of motion of VRR are mathematically similar to that of a point vortex system on the sphere \citep[see Equation~B84 in Appendix B in][]{Kocsis+Tremaine2015}. In this study we restrict our attention to an axisymmetric distribution of angular momentum vectors, and keep the simplifying assumption of a single-component system with the same semimajor axis, eccentricity, and mass for all objects \citet{Roupas+2017}.\footnote{However, note that it is straightforward to generalize the applied method for general multi-component systems with different couplings between different components as discussed in \citet{Roupas+2017}.} We may expect that the distribution is strongly modified by higher order multipoles in the case where the distribution forms a thin disk. But since the first order phase transition takes place between a thick disk and a spheroidal structure, one may expect a similar phenomenon to take place also in models with higher multipoles. Similarly, the negative temperature states are broadly distributed in inclination, which may indicate not to be greatly affected by high order moments. We verify these expectations quantitatively in this paper.

Important astronomical applications of VRR include the origin and evolution of a thin disk of massive stars in the Galactic center \citep[see in][]{Bartko:2008ad,Kocsis+Tremaine2011,0004-637X-783-2-131,2014ragt.conf...45H,2014IAUS..303..235H,2016MNRAS.458.4143S,2017MNRAS.465.1856S}. Furthermore VRR affects the distribution of putative stellar mass black holes in NSCs, which may represent important sources of gravitational waves (GWs) for existing and upcoming instruments: LIGO\footnote{\href{ligo.org}{www.ligo.org}}, VIRGO\footnote{\href{http://www.virgo-gw.eu/}{www.virgo-gw.eu}}, KARGA\footnote{\href{http://gwcenter.icrr.u-tokyo.ac.jp/en/}{www.gwcenter.icrr.u-tokyo.ac.jp}}, LISA\footnote{\href{https://www.elisascience.org/}{www.elisascience.org}}: \citep{OLeary:2008myb,2012ApJ...757...27A,Hoang:2017fvh,Bartos:2016dgn,McKernan:2013cha,McKernan:2014oxa,McKernan:2014hha,McKernan:2017umu}. The distribution of event rates as a function of mass and eccentricity may depend on the structure of the dynamical environments in which these GW sources form \citep{Gondan2017}. 

\section{Dynamics of VRR}
\subsection{Hamiltonian}

To examine VRR, the $N$-body gravitational Hamiltonian is averaged over the apisdal precession timescale. During this process, the Keplerian Hamiltonian of orbits around the SMBH are conserved for each orbit respectively; hence these conserved terms in the Hamiltonian may be dropped. This yields \citep{Kocsis+Tremaine2015}
\begin{align}\label{e:HVRR}
H_\text{VRR}
	=& -\frac{1}{2}\sum^{\substack{N}}_{\substack{i,j\\i\neq j}}\left\langle\frac{Gm_i m_j}{|\bm{r}_i(t)-\bm{r}_j(t')|}\right\rangle_{t,t'}=\nonumber\\
    =& -\frac{1}{2}\sum^{\substack{N}}_{\substack{i\neq j}}\sum^{\substack{\infty}}_{\substack{\ell=0}}\mathcal{J}_{ij\ell}P_\ell(\bm{\hat{L}}_i\cdot\bm{\hat{L}}_j),
\end{align}
where $i,j=1,...,N$ denote the bodies, $G$ is the gravitational constant and the distance $\bm{r}_i$ is measured from the SMBH, and $P_\ell(x)$ are Legendre polynomials. The dynamical variables are the directions of angular momentum vectors $\bm{\hat{L}}_i$. Coupling coefficients are generally given by \citet{Kocsis+Tremaine2015}, which are nonzero for even $\ell$ and depend on the mass, semi-major axes and eccentricity of the bodies.

In this paper, we examine the case of radially overlapping orbits. In this case the coupling coefficients follow asymptotically
\begin{equation}\label{eq:J}
\mathcal{J}_{ij\ell} = \mathcal{J}^\mathrm{a}_{ij}/\ell^2
\end{equation}
for large and even $\ell$, where $\mathcal{J}^\text{a}_{ij}$ does not depend on $\ell$ \citep[see Appendix B5 and B6 in][]{Kocsis+Tremaine2015}. For all odd $\ell$, $\mathcal{J}_{ij\ell}=0$. Note that this approximation to utilize this expression for all even $\ell$, which is complementary to the quadrupolar approximation, is expected to be sufficient to predict the  qualitative behavior of radially overlapping systems, as the net contribution of small $\ell$ terms is subdominant in this case, see Figure B1 in \citet{Kocsis+Tremaine2015}. The numerical value of $\mathcal{J}_{ij}^{\rm a}$ is not important, as we express the results in those units. The sum over $\ell$ asymptotically simplifies to
\begin{equation}
\begin{split}
H_\text{VRR}
    &\approx-\sum^{\substack{N}}_{\substack{i\neq j}}\frac{\mathcal{J}^\text{a}_{ij}}{2}\left[ 1-x_1-x_2+x^2_1\text{ln}\left(1+\frac{1}{x_1}\right)\right.\\
    &\,\quad+\left.x^2_2\text{ln}\left(1+\frac{1}{x_2}\right)\right],
\end{split}
\end{equation}
where we used the generator function of Legendre polynomials as discussed in Appendix B6 in \citet{Kocsis+Tremaine2015}, and introduced $x_{1,2}=(1\pm\bm{\hat{L}}_i\cdot\bm{\hat{L}}_j)/2)^{1/2}$.

\subsection{Mean-Field Theory}
We formulate the mean-field model for $N$ bodies with identical mass $m$, semimajor axis $a$, eccentricity $e$, and angular momentum magnitudes $l=m\sqrt{GM_\bullet a(1-e^2)}$. In this case $\mathcal{J}_{ij\ell}$ in Equations~\eqref{e:HVRR}--\eqref{eq:J} does not vary with $i$ and $j$. To separate the Hamiltonian, we expand $P_{\ell}(\bm{L}_i\cdot\bm{L}_j)$ using spherical harmonics \citep{jackson1975classical}, 
\begin{equation}
H_\text{VRR}
	= -\frac{1}{2}\sum^{\substack{N}}_{\substack{i\neq j}}\sum^{\substack{\infty}}_{\substack{\ell=0}}\mathcal{J}_{\ell}\frac{4\pi}{2\ell+1}\sum^{\substack{\ell}}_{\substack{m=-\ell}}Y^m_\ell(\bm{\hat{L}}_i){Y^m_\ell}^*(\bm{\hat{L}}_j).
\end{equation}
The system is characterized by the total number of bodies $N$, the total angular momentum $\bm{L}$, and the total orbit- and apsidal precession-averaged VRR energy $E$. We define the number of bodies with the direction of angular-momentum unit vector $\bm{n}$ oriented within an infinitesimal solid angle element $\mathrm{d}\Omega$ to be $f(\bm{n})\mathrm{d}\Omega$. The particle number, total angular momentum, and energy are respectively 
\begin{align}\label{eq:totalN}
N&=\int\mathrm{d}\Omega f(\bm{n})\,,\\
\bm{L}&=l\int\mathrm{d}\Omega f(\bm{n})\bm{n}\,,\\
E&= \frac{1}{2}N\langle\varepsilon\rangle.
\label{eq:total_energy}
\end{align}
where $\langle\varepsilon\rangle$ denotes the mean energy
\begin{align}\label{eq:epsilon}
\varepsilon(\bm{n}) 
	= - \mathcal{J}_{\ell}N\sum^{\substack{\infty}}_{\substack{\ell=0 }}\frac{4\pi}{2\ell+1}\sum^{\substack{\ell}}_{\substack{m=-\ell}}\langle{Y^{m}_\ell}^*\rangle Y^m_\ell(\bm{n}),
\end{align}
where the configuration average over all particles for any function $\bm{X}(\bm{n})$ is defined by
\begin{equation}.
\langle\bm{X}\rangle =\frac{1}{N}\int\mathrm{d}\Omega f(\mathbf{n})\bm{X}\,.
\end{equation}

\subsection{Statistical Equilibrium}
We determine the equilibrium distribution of $f(\bm{n})$ by extremizing the Boltzmann entropy
\begin{equation}
S=-k_B\int\mathrm{d}\Omega f(\bm{n})\ln f(\bm{n}),
\end{equation}
with respect to the equilibrium distribution for fixed total energy $E$, total angular momentum $\bm{L}$ and particle number $N$. This gives 
\begin{equation}
\delta S/k_B+\alpha\delta N-\beta\delta E+\bm{\gamma}\cdot\delta\bm{L}=0,
\end{equation}
where $\alpha$, $\beta$ and $\bm{\gamma}$ are the Lagrange multipliers of the constraints, Equations~(\ref{eq:totalN})--(\ref{eq:total_energy}). The resulting equilibrium distribution is \citep{Roupas+2017}
\begin{equation}\label{eq:f_eq}
f(\bm{n},\beta,\bm{\gamma},N)=N\frac{\text{e}^{-\beta\varepsilon(\bm{n})+l\bm{\gamma}\cdot\bm{n}}}{\int\mathrm{d}\Omega\text{e}^{-\beta\varepsilon(\bm{n})+l\bm{\gamma}\cdot\bm{n}}}.
\end{equation}
The order parameters satisfy the mean field self-consistency equation
\begin{equation}\label{eq:orderparameter}
\langle {Y^{m}_\ell}\rangle=\frac{1}{N}\int\mathrm{d}\Omega\,{Y^m_\ell}(\bm{n})f(\bm{n})
\end{equation}
where $f(\bm{n})$ is the equilibrium distribution function given by the order parameters by Equations~\eqref{eq:epsilon} and \eqref{eq:f_eq}. To avoid confusion, note that here $\langle {Y^{m}_\ell}\rangle$ in the left hand side of Equation~\eqref{eq:orderparameter} labels the unknown parameters that we determine by solving this equation, and the right hand side depends on this parameter through the equilibrium distribution $f(\bm{n})$ via $\varepsilon(\bm{n})$. Since the right hand side depends only on $\langle {Y^{m}_\ell}\rangle$ with even $\ell$, the $\langle {Y^{m}_{2\ell}}\rangle$ unknowns form a closed set of equations. Once determined, the distribution function follows from Equations~\eqref{eq:epsilon} and \eqref{eq:f_eq}. Note that given $\langle {Y^{m}_{2\ell}}\rangle$, the odd moments $\langle {Y^{m}_{2\ell+1}}\rangle$ may be evaluated directly from Equation~\eqref{eq:orderparameter}. They are also generally nonzero if $\bm{L}\neq 0$ since this implies that $\bm{\gamma}\neq 0$.

In equilibrium, the entropy and free energy are
\begin{align}\label{eq:entropy}
S(E,\bm{L})/k_B=2\beta E-\bm{\gamma}\cdot\bm{L}+N\text{ln}Z_0-N\text{ln}N,\\
\label{eq:free_energy}
F(T,\bm{L})=E-S/k_B=-E+\bm{\gamma}\cdot\bm{L}-N\text{ln}Z_0,
\end{align}
where $T=\beta^{-1}$ is the temperature and
\begin{equation}
Z_0=\int\mathrm{d}\Omega\text{e}^{-\beta\varepsilon+l\bm{\gamma}\cdot\bm{n}}.
\end{equation}

\subsection{Axisymmetric system}
In the rest of the paper we restrict our investigation for simplicity to an axisymmetric ($m=0$) radially overlapping system. \citet{Roupas+2017} found that for $\ell=2$, the maximum entropy configuration at any non-zero temperature is always axisymmetric, but at low temperatures non-axisymmetric metastable equilibria also exist which may be long-lived. Metastable equilibria may be studied by solving the nonlinear self-consistency system of equations~\eqref{eq:orderparameter} with $m\neq 0$, but this lies beyond the scope of this paper. We examine the canonical equilibrium distribution for fixed $(\beta,N,\bm{L})$. In this case, the self consistency equations for the order parameters are 
\begin{align}\label{eq:self_consistent_eq_system_axisym}
\langle Y^{0}_\ell\rangle&=\frac{\int\mathrm{d}\Omega\,\text{e}^{-\beta\varepsilon(\theta)+l\gamma \cos\theta}\,Y^{0}_\ell(\theta)}{\int\mathrm{d}\Omega'\,\text{e}^{-\beta\varepsilon(\theta')+l\gamma \cos \theta'}},\\
\frac{L}{Nl}&=\frac{\int\mathrm{d}\Omega\,\text{e}^{-\beta\varepsilon(\theta)+l\gamma \cos\theta}\,\cos\theta}{\int\mathrm{d}\Omega'\,\text{e}^{-\beta\varepsilon(\theta')+l\gamma \cos \theta'}},\\
\varepsilon(\theta)&=-\sum^{\substack{\infty}}_{\substack{\ell=0\\ \ell\text{ even}}}\frac{\mathcal{J}^\text{a} N}{\ell^2}\frac{4\pi}{2\ell+1}Y^0_\ell(\theta)\langle Y^{0}_\ell\rangle.\label{eq:self_consistent_eq_system_axisym2}
\end{align}
Entropy extrema have the property that $\bm{\gamma}$ is parallel with $\bm{L}$ which we choose along $\theta=0$ \citep[see Appendix B in][]{Roupas+2017}. 

By definition, the total angular momentum is bounded between $0\le L/(Nl)\le 1$, therefore, the order parameters are bounded\footnote{We use the definition $Y^0_\ell(x)=\sqrt{(2\ell+1)/4\pi}P_\ell(x)$, where $-1/2\le P_{2\ell}(x)\le 1$, and $\langle\cos\theta\rangle^n\le\langle\cos^n\theta\rangle$ for $n>1$ integer due to the Cauchy--Schwarz inequality.} between
\begin{equation}\label{eq:order_parameter_boundary}
\sqrt{\frac{2\ell+1}{4\pi}}P_\ell\left(L/(Nl)\right)\le\langle Y^0_{\ell}\rangle\le\sqrt{\frac{2\ell+1}{4\pi}}.
\end{equation}
The total energy is also bounded
\begin{equation}
\begin{split}
&-\frac{\pi^2}{48}\le\frac{E}{\mathcal{J}^\text{a}N^2}\le0,\hspace{3.5cm}\text{if }\frac{L}{Nl}\le\frac{L_*}{Nl},\\
&-\frac{\pi^2}{48}\le\frac{E}{\mathcal{J}^\text{a}N^2}\le-\sum_{\ell,\text{ even}}\frac{P_{\ell}\left(L/(Nl)\right)^2}{2\ell^2},\hspace{0.35cm}\text{if }\frac{L}{Nl}\geq\frac{L_*}{Nl},
\end{split}
\label{eq:energy_boundary}
\end{equation}
where we use Eqs.~\eqref{eq:total_energy} and \eqref{eq:order_parameter_boundary} and defined $L_*/(Nl)= 0.5488$. Note the upper bound on the VRR energy, which leads to the existence of negative temperature equilibria.

\subsection{Numerical method}
We determine the axisymmetric mean-field equilibrium distribution by numerically satisfying Eq.~\eqref{eq:self_consistent_eq_system_axisym} for fixed $\beta$ and $L/Nl$. To solve for $\langle Y^0_\ell\rangle$, we look for the zeros of the following function: 
\begin{equation}
F_{\ell}(\langle\bm{Y}\rangle)=\langle Y^0_\ell\rangle-\frac{\int\mathrm{d}\Omega\,{Y^0_\ell}(\bm{n})\text{e}^{-\beta\varepsilon(\theta)+l\gamma s}}{\int\mathrm{d}\tilde{\Omega}\,\text{e}^{-\beta \varepsilon(\tilde{\theta})+l\gamma\tilde{s}}}\,,
\end{equation}
where $\langle\bm{Y}\rangle $ is the set of all order parameters. Using the first order Newton's method, we Taylor's expand the equilibrium around some nearby order parameters to leading order
\begin{align}\label{eq:F(Y)}
F_\ell(\langle\bm{Y}^{[\rm eq]}\rangle) \approx& F_\ell(\langle\bm{Y}\rangle) + \sum_{\ell'}\left.\frac{\partial F_\ell(\langle\bm{Y}\rangle)}{\partial\langle Y^0_{\ell'}\rangle}\right|_{\bm{Y}}\nonumber\\
&\times\left[\langle Y^{0,\text{[eq]}}_{\ell'}\rangle-\langle Y^0_{\ell'}\rangle\right] 
\end{align}
The left hand side vanishes by definition at the equilibrium. The order parameter can be determined iteratively. In the  $M^\text{th}$ iteration, 
\begin{equation}\label{eq:Y(s,betaJN)}
Y^0_\ell{[M+1]}=Y^0_\ell{[M]}-\sum_{\ell'}\left(\frac{\partial F_\ell(\langle\bm{Y}\rangle)}{\partial\langle Y^0_{\ell'}\rangle}\right)^{-1}F_{\ell'}(\bm{Y}{[M]})\,,
\end{equation}
where ``$^{-1}$'' denotes the inverse matrix. The order parameters may be calculated with this method for any number of harmonics and tolerance. We carry out calculations by truncating the harmonics at $\ell_{\max}=10$, 38, and 58, respectively. Deviations for $\ell_{\max}=10$ and 38 from 58 is of order $\mathcal{O}(10^{-3})$ and $\mathcal{O}(10^{-6})$, respectively, for $\beta\mathcal{J}^\text{a}N=50$ and $L/Nl=0.1$.

\begin{figure*}
\includegraphics[width=0.5\linewidth]{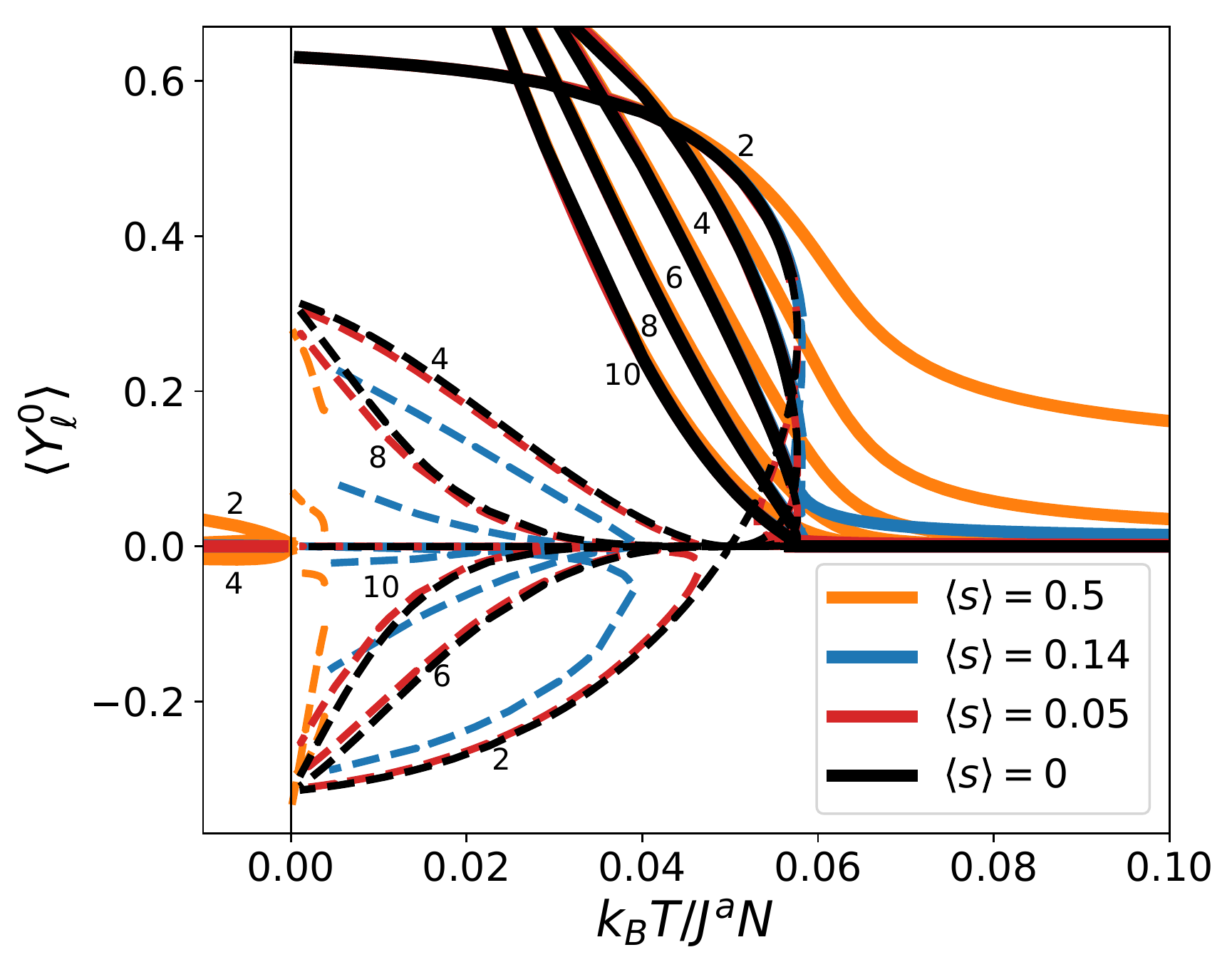}
\includegraphics[width=0.494\linewidth]{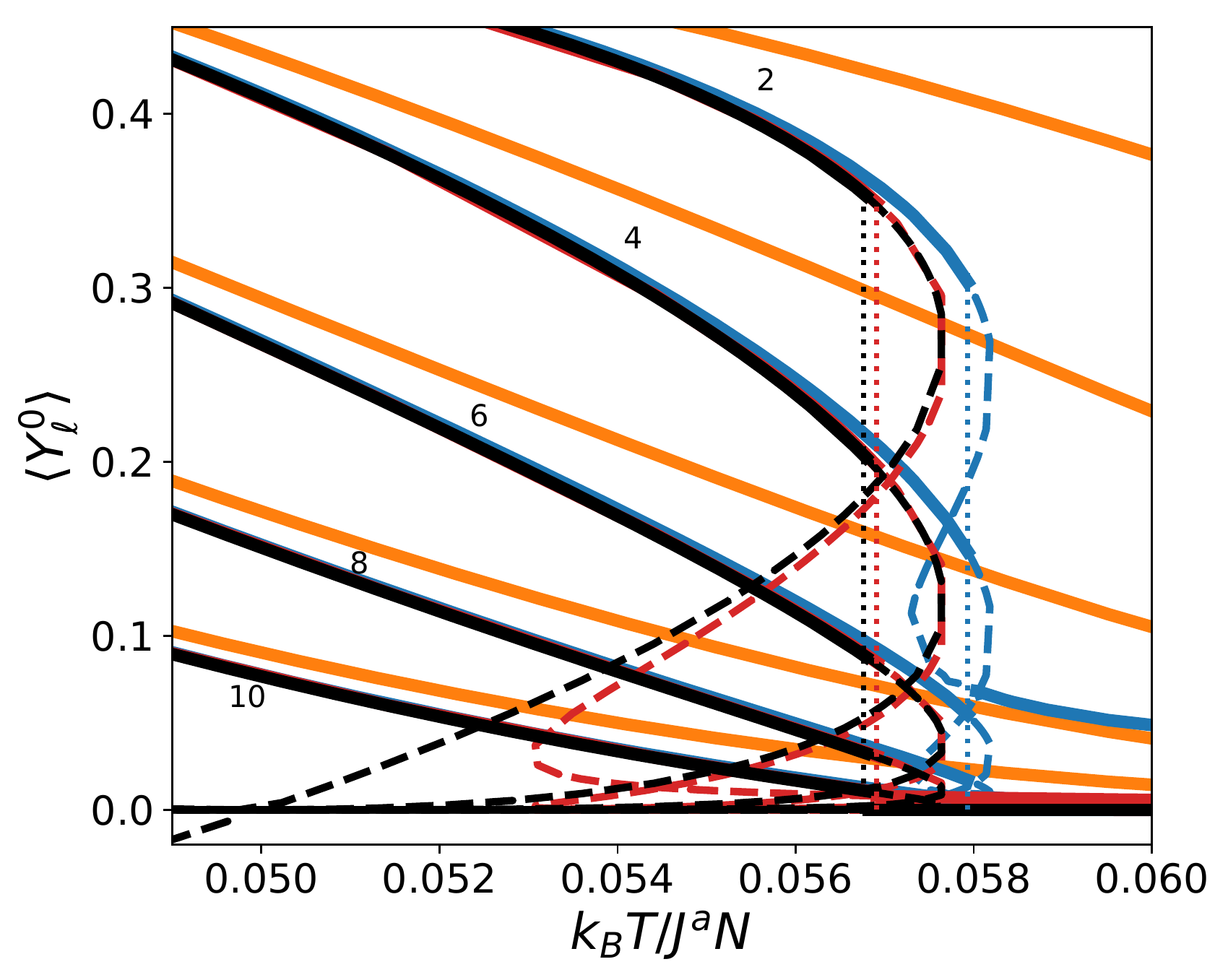}
\caption{\label{fig:Q_kTperJN_Nharm_6_final} The first five even multipole moments (order parameters) vs. temperature for different total angular momenta, $L$, labeled using $\langle s\rangle\equiv L/(Nl)$. Here $s=\cos\theta$ is the polar angle of angular momentum vectors in spherical coordinates aligned with $\bm{L}$. The right panel is a zoom-in of the left panel to highlight the phase transition in the canonical ensemble. Solid (dashed) lines represent stable (meta and unstable) equilibria in the canonical ensemble. For $L/(Nl)<0.188$ the order parameters' change is discontinuous, which results in a first order phase transition in the canonical ensemble between the ordered and disordered) phase at low and high temperatures, respectively.
}
\end{figure*}

\begin{figure*}
\includegraphics[width=0.5\linewidth]{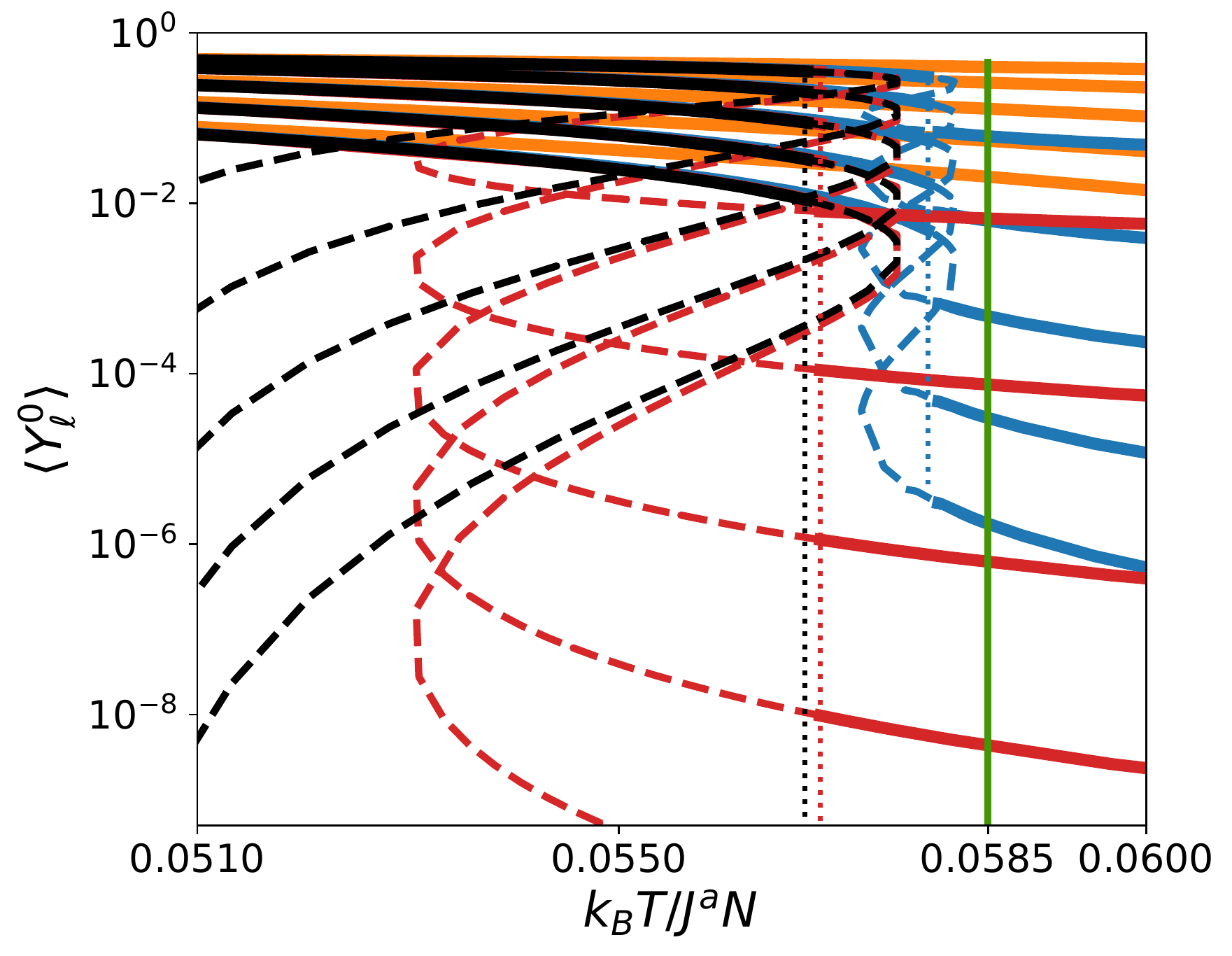}
\includegraphics[width=0.49\linewidth]{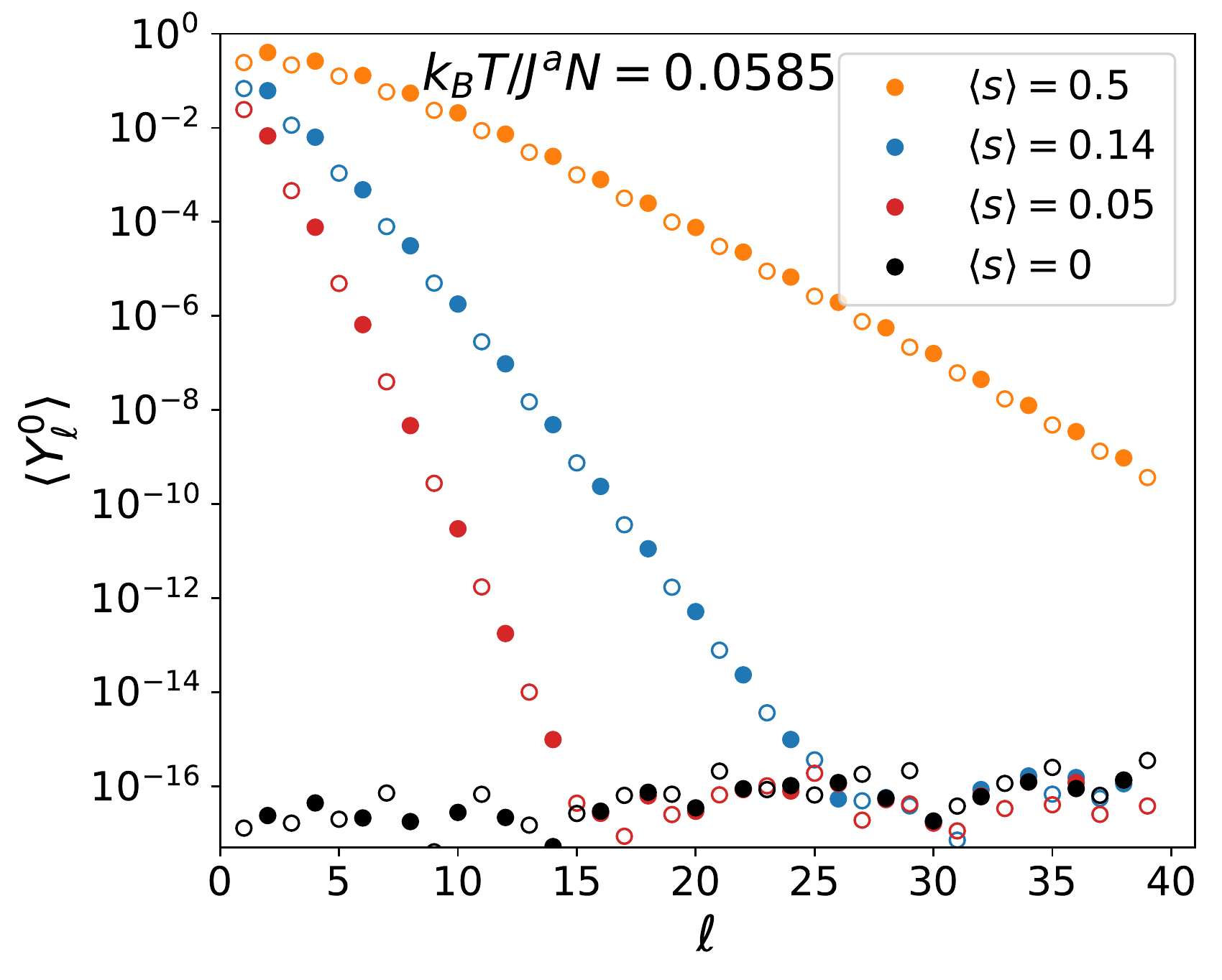}
\caption{\label{fig:logQ_kTperJN_Nharm_6_final} \textit{Left panel:} same as the right panel of Figure~\ref{fig:Q_kTperJN_Nharm_6_final}, showing the multipole moments with $\ell=2$, 4, 6, 8, and $10$ vs. temperature for different total angular momenta but on a logarithmic scale. The phase transition temperature is highlighted with vertical dotted lines. \textit{Right panel:} the order parameters $\langle Y_\ell^0\rangle$ as a function of $\ell$ in the disordered phase at the temperature marked by a green vertical line in the left panel. Even and odd $\ell$ are shown with filled and empty circles. Multipoles are exponentially suppressed in the disordered state as a function of the harmonic index $\ell$. The saturation below $10^{-15}$ is an artifact of numerical errors.
}
\end{figure*}

\section{Results}
We determine the numerical solution for the one-component axisymmetric system at fixed $(\beta,N,L)$. The results are qualitatively very similar to that for the quadrupole interaction \citet{Roupas+2017}. We direct the reader for that paper for extensive discussions. To compare the two results, the following conversions are needed: $J_\text{Roupas}=2\mathcal{J}^\text{a}/3$ and $\langle q\rangle_\text{Roupas}=\sqrt{16\pi/45}\langle Y^0_2\rangle$.

\subsection{Order parameters and phase transition}
Figure~\ref{fig:Q_kTperJN_Nharm_6_final} and \ref{fig:logQ_kTperJN_Nharm_6_final} show the order parameters for axisymmetric one-component systems, $\langle Y_{\ell}^{0}\rangle$ for even $\ell$, which solve Eqs.~\eqref{eq:self_consistent_eq_system_axisym}--\eqref{eq:self_consistent_eq_system_axisym2} as a  function of temperature, assuming a truncation of the harmonics at $\ell_{\max}=10$. Different colors show different fixed total angular momentum per particle $\langle s\rangle=L/(Nl)$, where we introduced $s=\cos\theta$. The $\ell$ harmonics are labeled with small numbers. Solid and dashed lines belong to stable and metastable/unstable equilibria in the canonical ensemble.\footnote{Due to the non-additivity of the system, the stability of equilibria are different in the canonical and the microcanonical ensembles, a phenomenon known as ensemble inequivalence \citep{campa2014physics,Roupas+2017}. For VRR, there is no phase transition in the microcanonical ensemble, where the system changes continuously from the flattened state to a nearly spherical state along the S-shaped curves shown in the right panel of Figure~\ref{fig:Q_kTperJN_Nharm_6_final} for all $\ell$. }
The right panel is a zoom-in of the left panel, which highlights the phase transition shown with dotted lines.
At low temperatures, the system forms an ordered phase, which is analogous to the \textit{nematic phase} of liquid crystals\footnote{The quadrupole approximation of VRR Hamiltonian is similar to the Maier--Saupe model of liquid crystals, where the ordered phase is called nematic \citep[see][for a discussion]{Roupas+2017}.}. The high temperature phase represents a nearly spherical disordered phase. We find that if the total angular momentum is less than a \textit{critical total angular momentum} $L_{\rm C}/(Nl)=0.188$, there is a discontinuous change in the order parameters, i.e. the multipole moments $\langle Y^0_{\ell}\rangle$, at the phase transition temperature. The phase transition in this case is first order. The phase transition becomes second order at $L=L_{\rm C}$ and a \textit{critical temperature} $k_{\rm B}T_{\rm C}/(\mathcal{J}^{\rm a}N)=0.05883$ and there is a continuous series of equilibria at higher $L$. Figure \ref{fig:logQ_kTperJN_Nharm_6_final} and the right panel of Figure \ref{fig:Q_kTperJN_Nharm_6_final} show that the absolute change of the order parameters at the phase transition is most prominent for the quadrupole $\ell=2$. The phase transition erases the high order multipoles (i.e. small scale anisotropies) much more than the low order multipoles approximately exponentially with $\langle Y_{\ell}\rangle\propto\exp(-\alpha\ell)$, where $1\lesssim\alpha\lesssim 3$, higher $\alpha$ values correspond to higher temperatures and lower total angular momentum. 

\begin{figure*}
\includegraphics[width=0.33\linewidth]{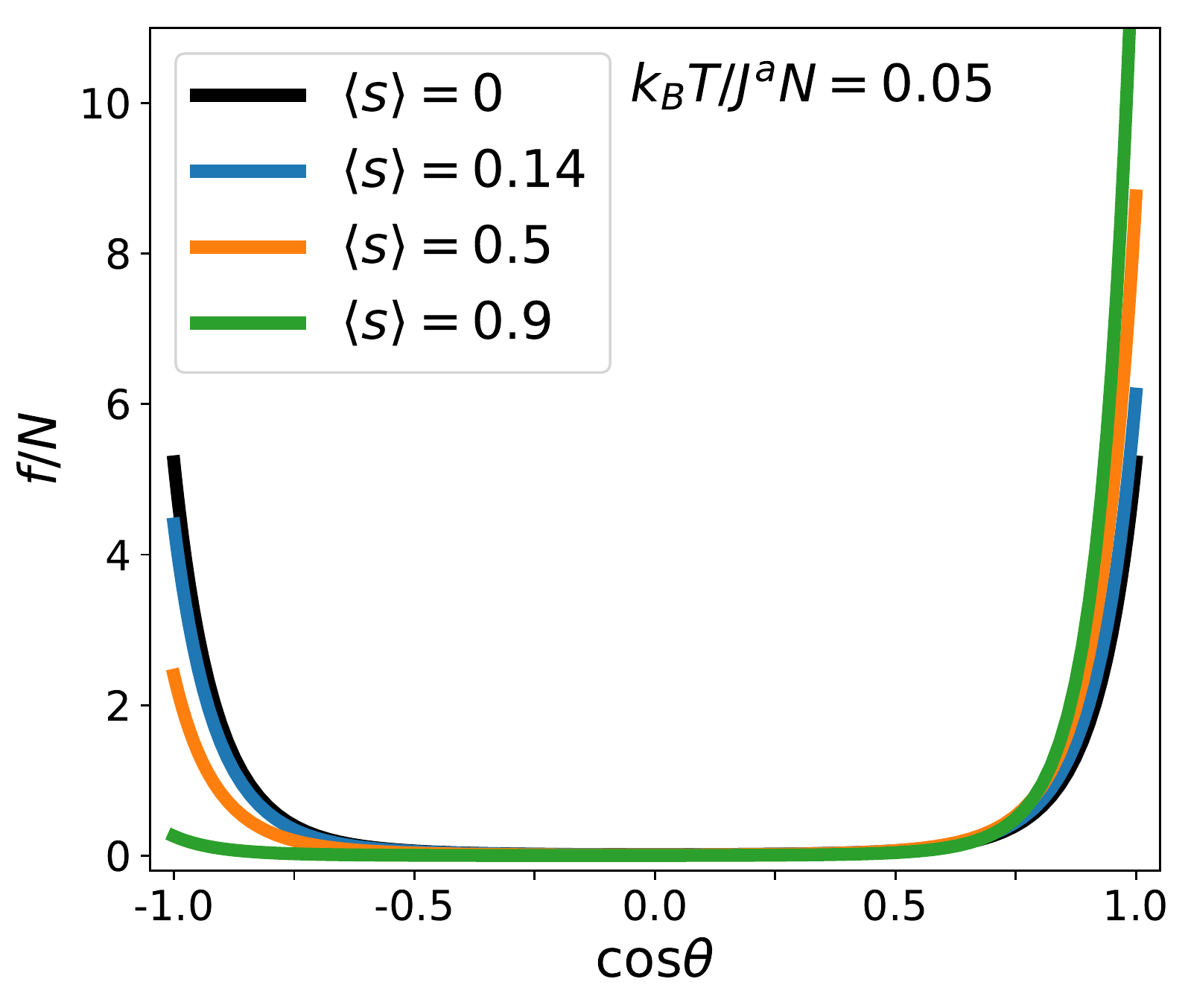}
\includegraphics[width=0.323\linewidth]{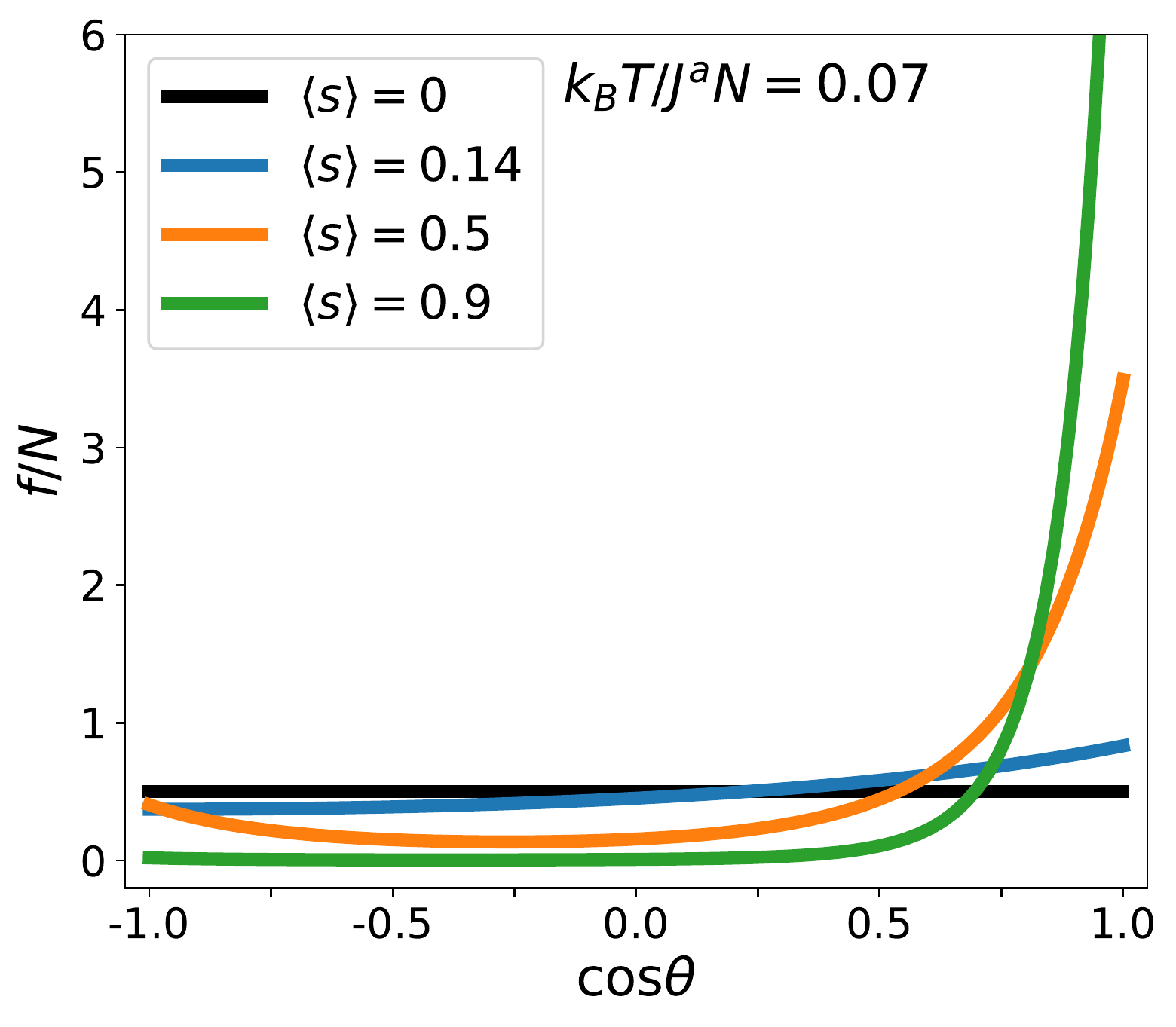}
\includegraphics[width=0.323\linewidth]{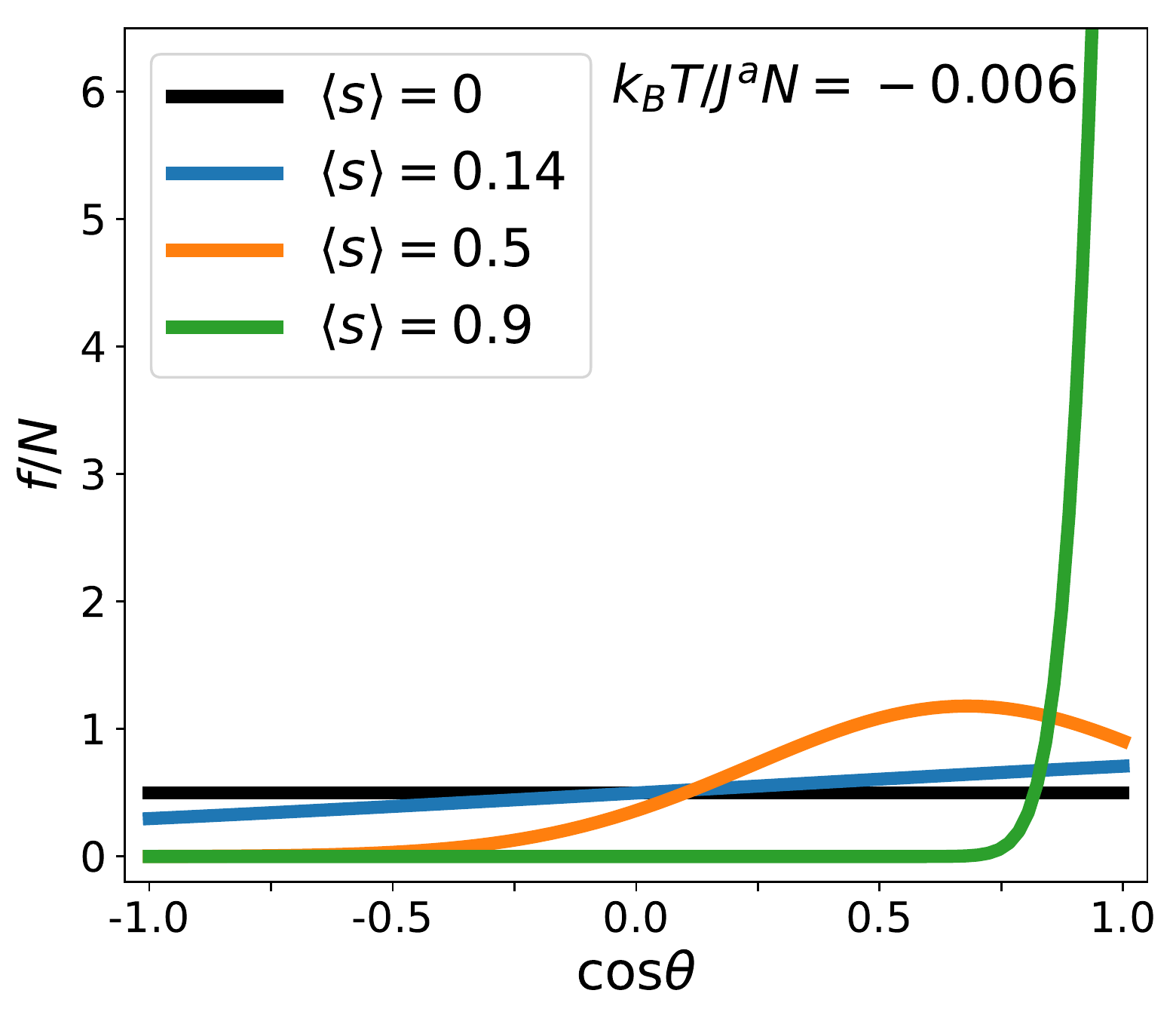}
\caption{\label{fig:feq_Nharm_6} Angular momentum distribution of stable equilibria for different total angular momenta shown with different colors. The left panel shows the ordered phase at $k_\mathrm{B}T/\mathcal{J}^\mathrm{a}N=0.05$. The middle and right panels show states in the disordered phase at positive and negative temperatures, respectively, as labeled. 
}
\end{figure*}

The highest entropy configuration satisfies $\beta=0$, $T\rightarrow \pm\infty$. In this case, the angular momentum vector distribution function is identical to that of \citet{Roupas+2017}, $f(\bm{n})\propto e^{l\gamma \cos\theta}$, and so the order parameters are
\begin{equation}
\langle Y_{\ell}^0\rangle = \sqrt{\frac{2\ell+1}{4\pi}}\frac{ \int_{-1}^1 \mathrm{d} s \, P_{\ell}(s) e^{l\gamma s}}{\int_{-1}^1 \mathrm{d} s \, e^{l\gamma s}}\qquad (T\rightarrow\pm\infty)\,,
\end{equation}
where $P_{\ell}(s)$ are Legendre polynomials and $l\gamma$ is given by
\begin{equation}
\frac{L}{Nl} = \frac{ \int_{-1}^1 \mathrm{d}s \, s e^{l\gamma s}}{\int_{-1}^1 \mathrm{d} s \, e^{l\gamma s}} = \coth (l\gamma) - \frac{1}{l\gamma}\qquad (T\rightarrow\pm\infty)\,.
\end{equation}
The moments are exponentially suppressed as a function of $\ell$ for $\ell\gtrsim(l\gamma)^{1/2}$.
Note that while the even moments fully parameterize the distribution function (Equations~\ref{eq:epsilon} and \ref{eq:f_eq}), the odd $\ell$ moments are also generally non-zero.

Figure~\ref{fig:feq_Nharm_6} shows the angular momentum vector distribution for axisymmetric equilibria at various temperatures. It is sharply peaked near the $\pm z-$axis in the low-temperature ordered phase ($k_\mathrm{B}T/\mathcal{J}^{\mathrm{a}}N=0.05$, left panel). This configuration corresponds to a counter-rotating thin disk in physical space, in which the ratio of bodies orbiting in the two senses is set by the value of the total angular momentum. The angular momentum distribution in the high temperature disordered phase ($k_\mathrm{B}T/\mathcal{J}^{\mathrm{a}}N=0.07$) is shown in the middle panel. The distinction between the two cases is clear for low total angular momentum which exhibits a phase transition. The right panel shows negative temperature equilibria which is similar to the positive temperature disordered phase. 

\subsection{Thermodynamical Properties and Instability}\label{sec:stability}
To explore the phase transition and the stability of states, we calculate the thermodynamical properties of canonical equilibria. 

\begin{figure*}
\includegraphics[height=0.4\linewidth]{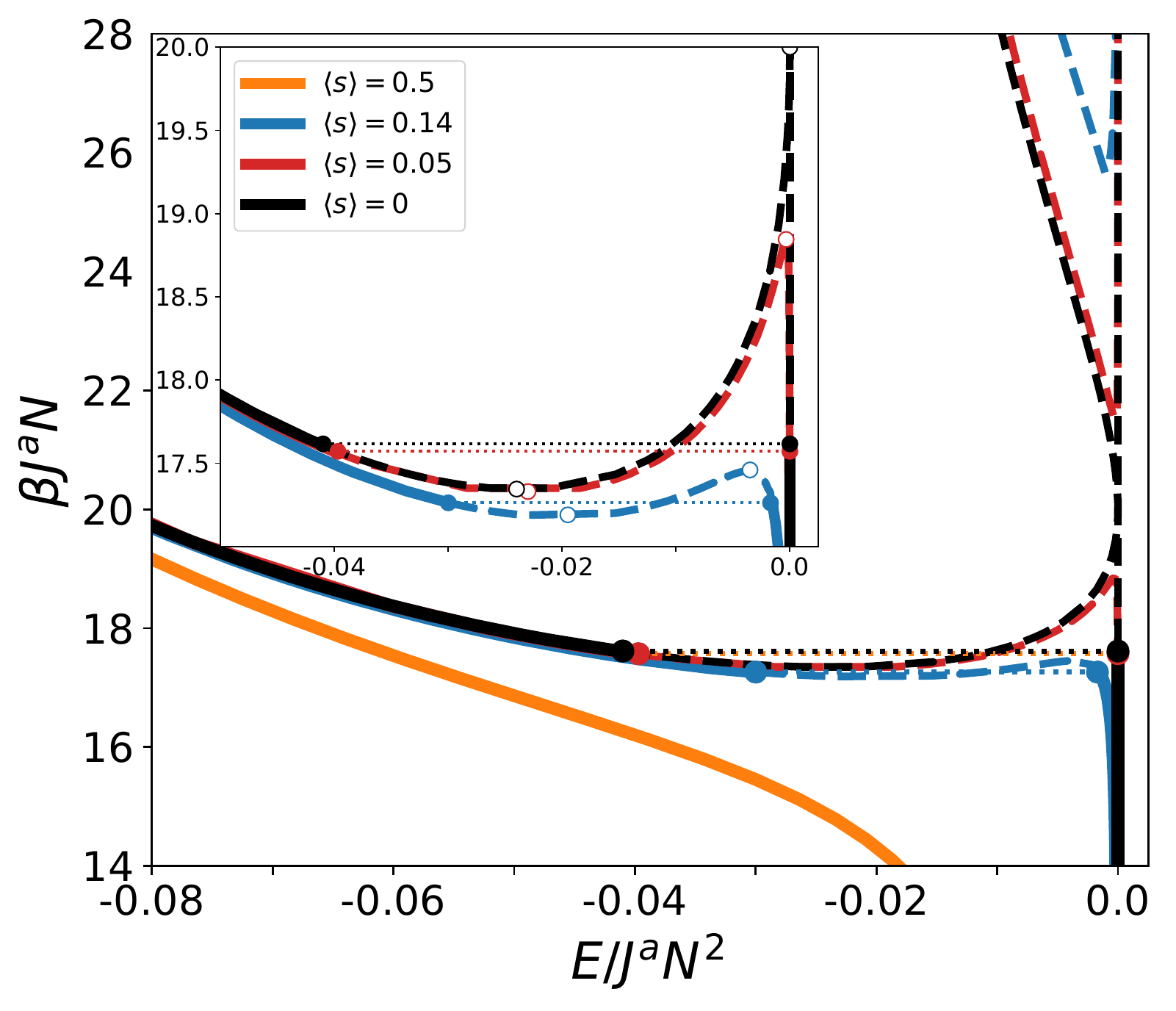}
\includegraphics[height=0.4\linewidth]{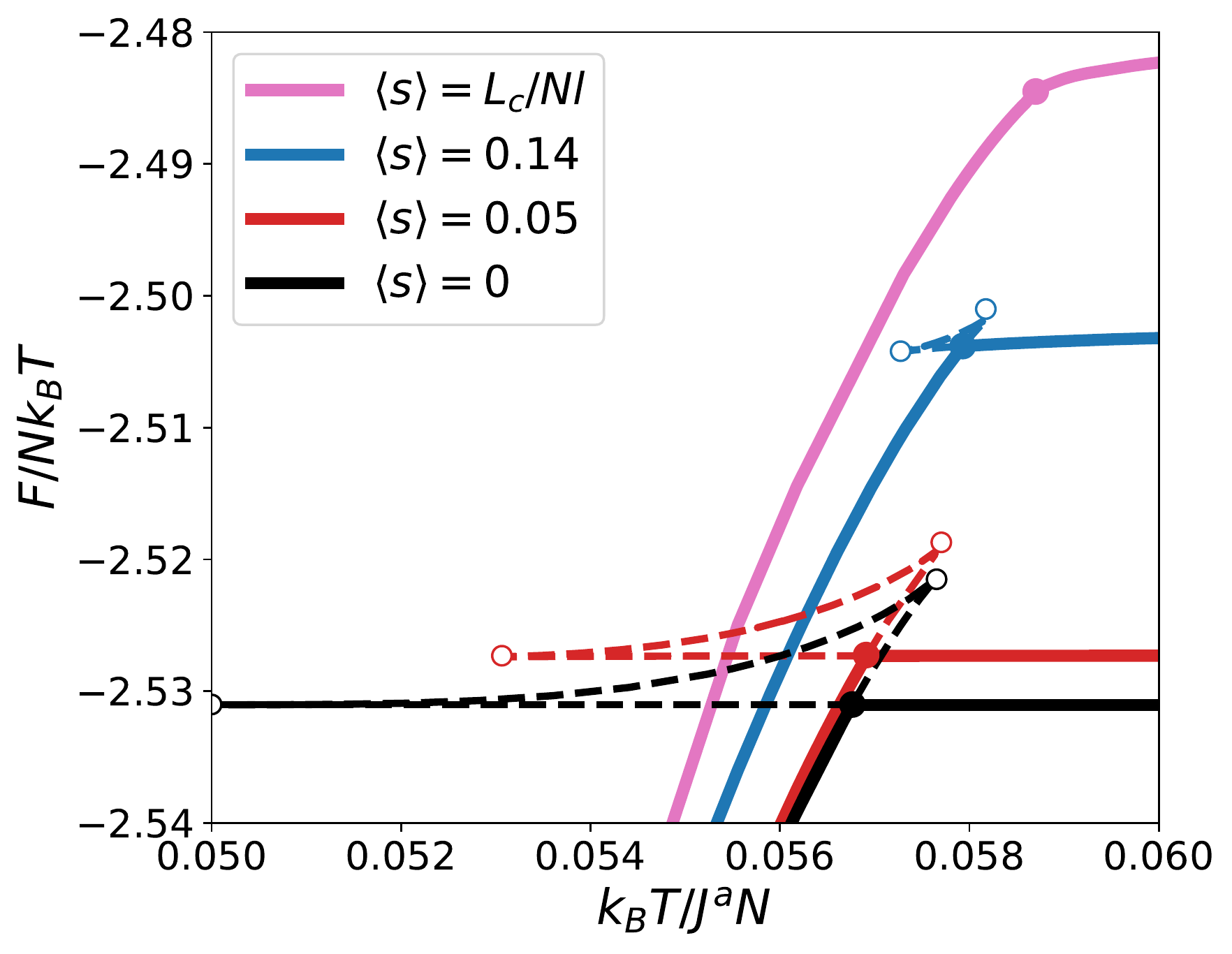}
\caption{\label{fig:EperJN2_betaJN_Nharm_6} \textit{Left:} Inverse temperature as function of total energy for different total angular momenta, plotted with different colors. First-order phase transition is characterized by the discontinuous change in total energy between the marked filled circles. The inset highlights the phase transition. \textit{Right:} Canonical free energy as a function of temperature for different total angular momenta. Solid lines show stable equilibria (free energy minimum at fixed temperature). The curved triangle near the phase transition (filled circles) represent metastable superheated/supercooled and unstable equilibria. }
\end{figure*}

The left panel of Figure~\ref{fig:EperJN2_betaJN_Nharm_6} shows the \textit{caloric curve}, i.e. the inverse temperature as a function of total energy from Eq.~\eqref{eq:total_energy}. There is a discontinuous change in energy between the marked points that characterize a first-order phase transition, which is the \textit{latent heat}. The temperature of the phase transition is shown with dotted lines. Between the filled and empty circles (local minima and maxima) the system is in metastable equilibria i.e. superheated or supercooled.
The right panel shows the free energy (Eq.~\eqref{eq:free_energy}) as a function of temperature. The free energy for a fixed temperature is minimized for stable equilibria (solid lines). As the order parameters are changed continuously along the series of equilibria with fixed $\langle s\rangle$ from the low energy ordered states (see Fig.~\ref{fig:Q_kTperJN_Nharm_6_final}), the free energy increases with temperature, then goes around a triangular shape. The phase transition takes place at the lower right vertex where the solid curve has a discontinuous derivative. The temperature and free energy are constant along the phase transition. We find that the series of equilibria shown in Figure~\ref{fig:EperJN2_betaJN_Nharm_6} is qualitatively very similar to Figure 11 in \citet{Roupas+2017}, see further discussion about the first order phase transition therein. 

\begin{figure*}
\includegraphics[height=0.38\linewidth]{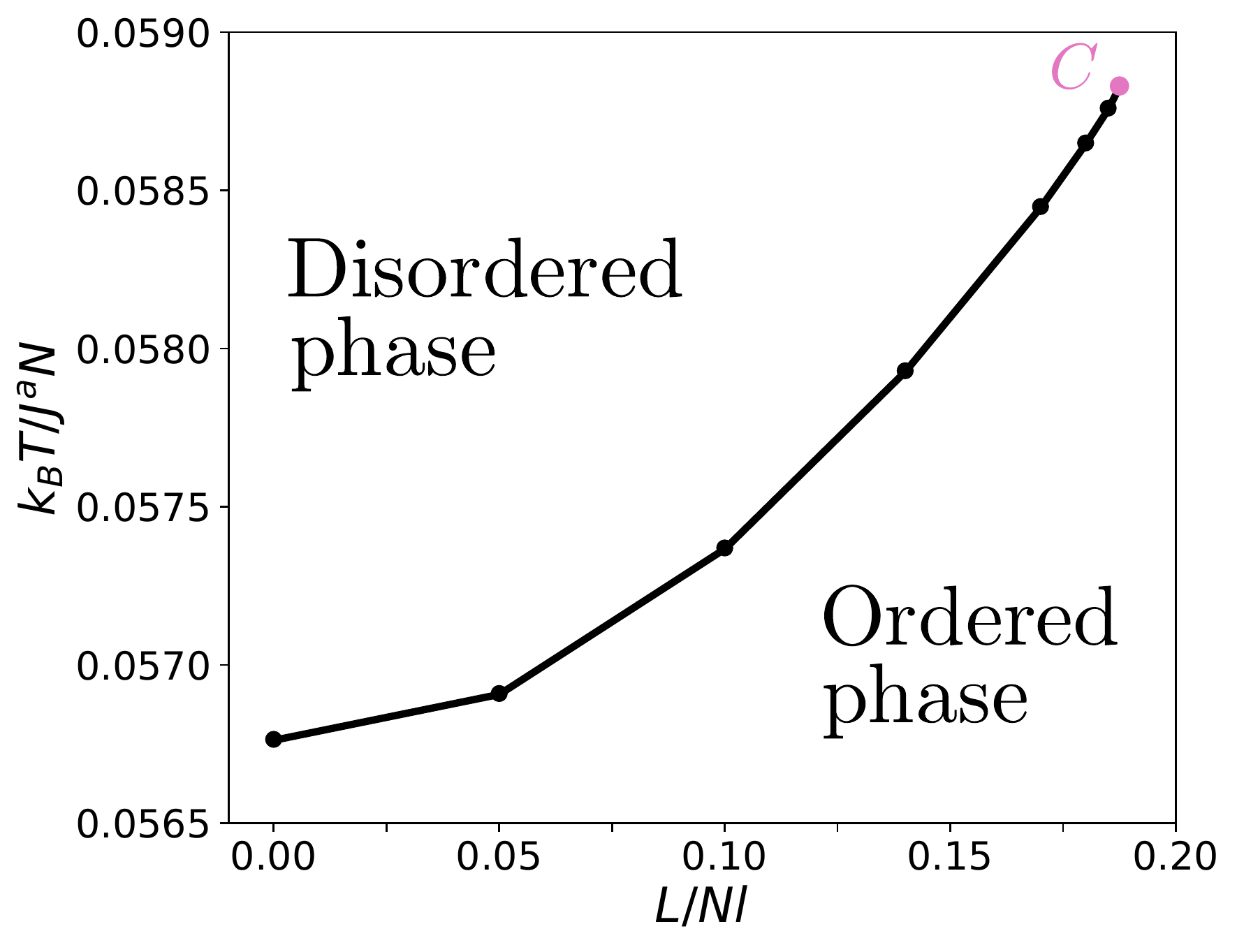}
\includegraphics[height=0.38\linewidth]{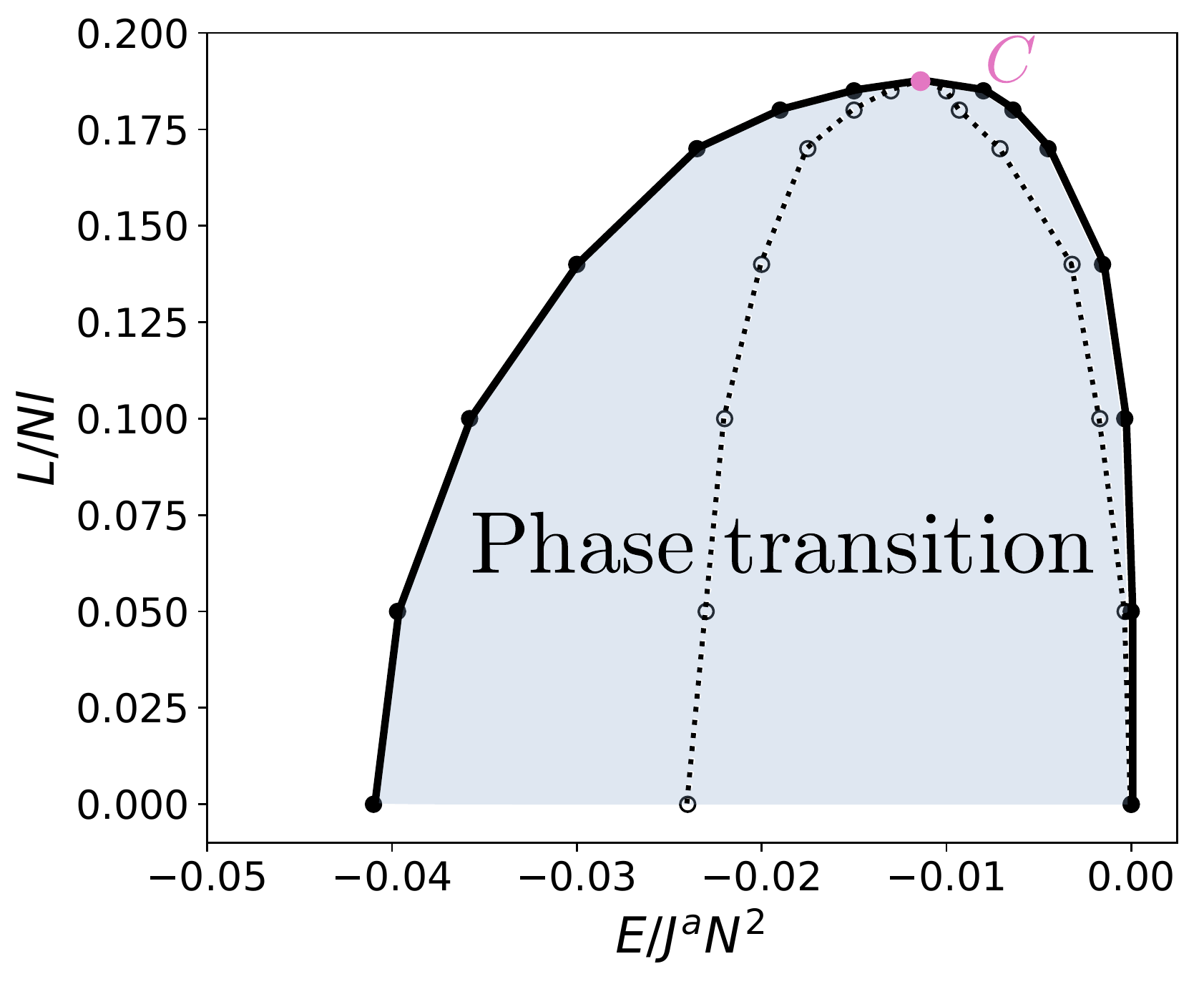}
\caption{\label{fig:phase diagram} The phase diagram of VRR. The ordered and disordered phases appear respectively at temperatures smaller and larger than the critical curve shown in the left panel below a critical angular momentum $L_{\rm C}/Nl=0.188$. There is a first order phase transition between the two phases in the canonical ensemble. The phase transition is second ordered at the critical point $C$, and there are continuous series of equilibria for $L>L_{\rm C}$. In the energy-angular momentum plane (right panel), the two phases are to the left and right of the phase transition region. The region enclosed by the dotted line is unstable in the canonical ensemble, and the regions outside of the dotted line but within the phase transition region represent supercooled and superheated metastable states. The calculations were evaluated at the large dots, which are connected by straight lines for clarity.
}
\end{figure*}

Figure \ref{fig:phase diagram} shows the phase diagram of the system.

\subsection{Negative Temperature Equilibria}
\begin{figure}
\includegraphics[width=\linewidth]{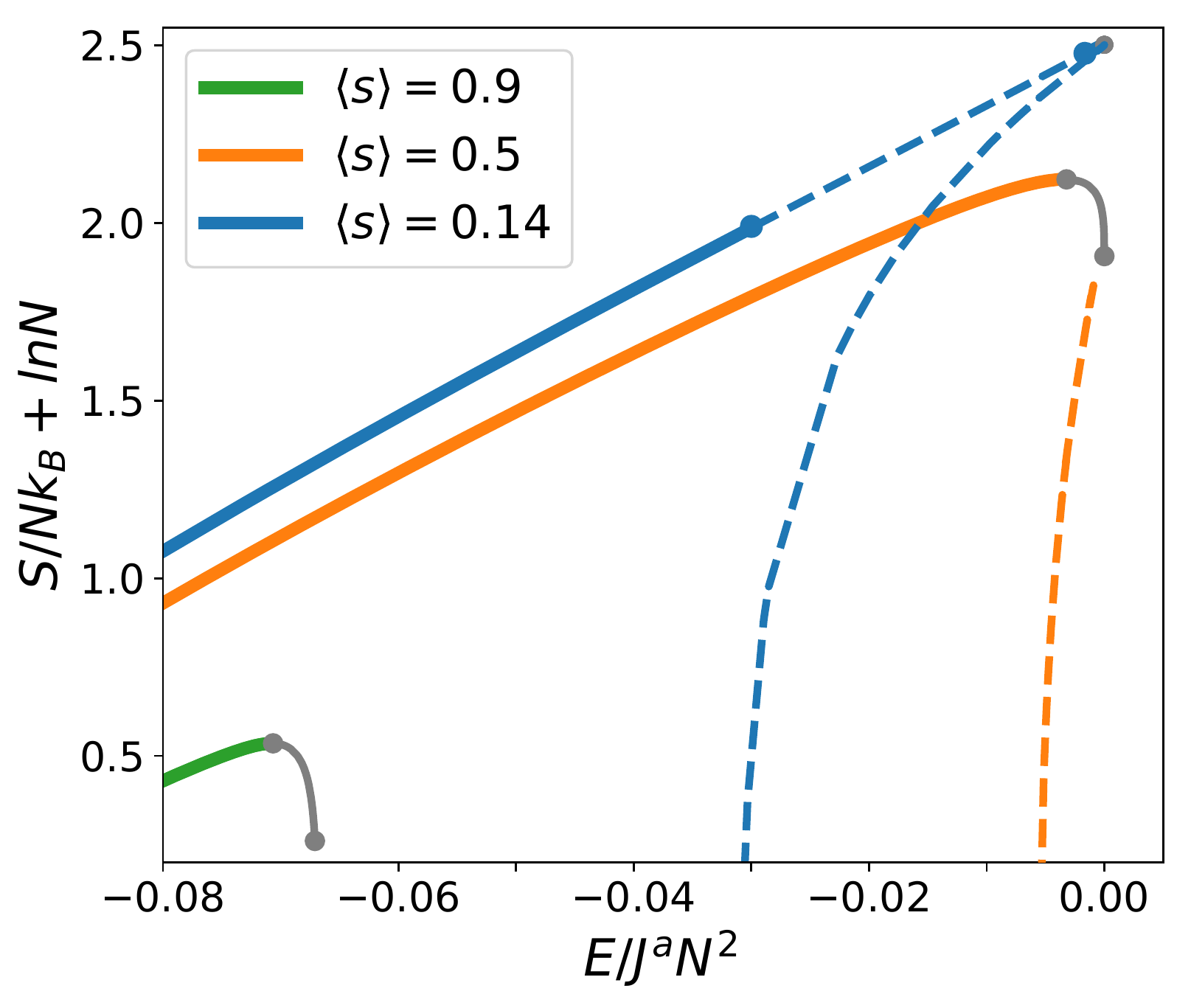}
\caption{\label{fig:SperkN_plus_lnN_Nharm_6} The entropy as function of total energy for different total angular momenta. Negative temperature equilibria are the segments with decreasing $S(E)$ between the point of maximum entropy and maximum energy. Solid/dashed curves show stable and unstable equilibria in the canonical ensemble. For the blue curve the dashed curve between the blue circles is the phase transition region in the canonical ensemble, where $S(E)$ is convex.
}
\end{figure}

To understand the origin of negative temperature equilibria in Figures~\ref{fig:Q_kTperJN_Nharm_6_final}~and~\ref{fig:feq_Nharm_6}, Figure~\ref{fig:SperkN_plus_lnN_Nharm_6} shows the entropy as function of total energy. By definition, entropy is a decreasing function of energy at negative temperatures, marked with solid gray lines. The origin of the negative temperature states stems from the fact that the VRR energy and the entropy are bounded from above\footnote{generally because the phase space is compact, and see Eq.~\eqref{eq:energy_boundary}}, which leads to a decreasing entropy at energies higher than that of the maximum entropy state. A first order phase transition appears where the $S(E)$ function is convex. A careful analysis shows that this is indeed the case between the large dots connected by a dashed curve on the blue curve. Note that for this series of equilibria, the negative temperature states subtend a rather small range of energies above the phase transition. Negative temperature states appear more prominently at higher total angular momenta or high $\langle s\rangle$, which do not admit a phase transition. The negative temperature region is always the highest entropy for fixed energy, and therefore stable in the microcanonical ensemble. Where $S(E)$ is concave, showing that there is no phase transition at negative temperatures. The qualitative shape of $S(E)$ is very similar in the quadrupole interaction approximation, discussed in detail in \citet{Roupas+2017}. 

The angular momentum vector distribution of negative temperature states is approximately isotropic for small total angular momentum $L/(Nl)$. The $\langle Y^{0}_{\ell}\rangle$ order parameters have alternating signs as a function of even $\ell$ for negative temperatures (Figure \ref{fig:Q_kTperJN_Nharm_6_final}), they are continuous functions of $\beta=1/T$ over $\beta=0$ and $\langle Y^{0}_{\ell}\rangle=0$ for even $\ell$ for $T=0$. Figures \ref{fig:Q_kTperJN_Nharm_6_final} and \ref{fig:logQ_kTperJN_Nharm_6_final} show that $\langle Y^{0}_{\ell}\rangle$ are highly suppressed for increasing even $|\ell|$ at negative temperatures.

\section{Discussion}\label{sec:discussion}
In this paper we have examined the statistical equilibrium distribution of orbital planes of gravitating bodies orbiting around a central massive object. We have considered a one-component system, in which all bodies have approximately equal semimajor axes and eccentricities, but relaxed the quadrupole interaction approximation applied recently in \citet{Roupas+2017}. The higher multipole moments modify the interaction significantly for radially overlapping orbits, particularly for a low mutual inclination \citep{Kocsis+Tremaine2015}. While the interaction Hamiltonian in the quadrupole approximation is that of liquid crystals \citep[see][for detailed discussions]{Roupas+2017}, for the case of overlapping orbits the VRR interaction resembles a vortex system  \citep[see Equation~B84 in Appendix B in][]{Kocsis+Tremaine2015}. Despite these differences, we found no qualitative difference in their statistical physics behavior between the case limited to the quadrupole approximation and that including high order harmonics. All of the physical features seen in that model \citep{Roupas+2017} is also present in the complementary model for overlapping orbits
in which we extrapolated the dominant asymptotic contribution of higher harmonics $\ell\rightarrow\infty$ to all even $\ell$: 
\begin{itemize}
\item Low-temperature axisymmetric equilibria resemble thin counterrotating disks in physical space in which the bodies may orbit in either sense, which is set by the total angular momentum.
\item Below a critical total angular momentum, $L_{\rm C}=0.188\,Nl$ the system exhibits a first order phase transition in the canonical ensemble between the ordered phase which resembles a thick disk and the disordered phase which resembles a spheroidal distribution.
\item The disordered phase for $L<L_{\rm C}$ is not completely isotropic if the total angular momentum is nonzero, but the multipole moments are exponentially suppressed as a function of the harmonic number $\ell$. Thus, the phase transition erases the small-scale features most efficiently.
\item The system admits a maximum energy and maximum entropy which leads to the existence of negative temperature states.
\end{itemize}
Possible differences may occur at low temperatures for anisotropic states. For example various arrangements of razor thin disks are expected to be stable during VRR discussed in \citet{Roupas+2017}. However, since the high temperature states generally do not have sharp density peaks in angular momentum direction space, we expect their behavior to be described well by the quadrupole approximation, at least for a one-component systems. This expectation may not hold for multicomponent systems, where heavier objects may be expected to form much thinner disks \citep{Roupas+2017}. A study of anisotropic or multicomponent equilibria lies beyond the scope of this paper. 

This result provides a point of comparison for interpreting the results of more detailed investigations on the equilibrium configurations of complex astrophysical systems with a distribution of overlapping orbits with different mass, eccentricity, and semimajor axis. It may have applications in explaining the origin of a thin disk of massive stars in the Galactic center \citep{Bartko:2008ad,Kocsis+Tremaine2011,0004-637X-783-2-131,2014ragt.conf...45H,2014IAUS..303..235H,2018arXiv180203027P,2018arXiv180200012P} and the distribution of possible putative stellar mass black holes lurking in nuclear star clusters. These elusive objects may be possibly observed by future X-ray observations including Chandra, XMM-Newton, and NuSTAR \citep{2013PhRvL.110v1102B}, and may represent important sources of mergers for GW detectors including LIGO, VIRGO, KARGA, and LISA \citep{OLeary:2008myb,2012ApJ...757...27A,Hoang:2017fvh,Bartos:2016dgn,McKernan:2013cha,McKernan:2014oxa,McKernan:2014hha,McKernan:2017umu}. The distribution of GW event rates may carry information on the structure of the dynamical environments in which these GW sources form \citep{Gondan2017}. 

We conclude that for one-component axisymmetric systems, the thermodynamics of VRR is qualitatively well described by the quadrupole approximation.

\acknowledgments{
This project has received funding from the European Research Council (ERC) under the European Union's Horizon 2020 research and innovation programme ERC-2014-STG under grant agreement No 638435 (GalNUC) and by the Hungarian National Research, Development, and Innovation Office grant NKFIH KH-125675. The calculations were carried out on the NIIF HPC cluster at the
 University of Debrecen, Hungary.
}

\bibliographystyle{yahapj}
\bibliography{main}
 
\end{document}